\documentclass[onecolumn]{aastex631}
\usepackage{natbib}
\usepackage{float}
\usepackage{textgreek}
\usepackage{graphicx}
\usepackage{array}
\usepackage{booktabs}
\usepackage{mathrsfs}
\usepackage{amsmath}
\usepackage{xcolor}
\usepackage{gensymb}

\def\agile{\textit{AGILE} }
\def\agilep{{\textit{AGILE}}}

\shorttitle{AGILE observations of a sample of repeating Fast Radio Burst sources}
\shortauthors{Casentini et al.}

\graphicspath{{./}{figures/}}

\begin{document}

\title{AGILE observations of a sample of repeating Fast Radio Burst sources}

\author[0000-0001-8100-0579]{Claudio Casentini}
\email{claudio.casentini@inaf.it}
\affiliation{INAF/IAPS, via del Fosso del Cavaliere 100, I-00133 Roma (RM), Italy}
\affiliation{INFN Tor Vergata, via della Ricerca Scientifica, I-00133 Roma (RM), Italy}

\author[0000-0003-3455-5082]{Francesco Verrecchia}
\email{francesco.verrecchia@inaf.it}
\affiliation{SSDC/ASI, via del Politecnico snc, I-00133 Roma (RM), Italy}
\affiliation{INAF/OAR, via Frascati 33, I-00078 Monte Porzio Catone (RM), Italy}

\author[0000-0003-2893-1459]{Marco Tavani}
\affiliation{INAF/IAPS, via del Fosso del Cavaliere 100, I-00133 Roma (RM), Italy}

\author[0000-0001-7397-8091]{Maura Pilia}
\affiliation{INAF/OAC, via della Scienza 5, I-09047, Selargius (CA), Italy}

\author[0000-0001-6897-5996]{Luigi Pacciani}
\affiliation{INAF/IAPS, via del Fosso del Cavaliere 100, I-00133 Roma (RM), Italy}

\accepted{by ApJ on March 7, 2025}

\begin{abstract}
Fast Radio Bursts are millisecond-duration bursts originating from distant sources. They are classified into two categories: non-repeating FRBs, which manifest as singular events, and repeating FRBs, which emit multiple bursts over time.

In this work, we report a search for X- and Gamma-ray counterparts to a selected sample of R-FRBs using data from the \agile
satellite. The sample focused on sources with an excess dispersion measure below $300 \, {\rm pc \, cm^{-3}}$. The analysis focused on the bursts covered by AGILE Mini-Calorimeter high resolution data.
No astrophysical signals were identified, and we derived upper limits on the flux above 400 keV for the associated sources adopting a spectral magnetar model, one of the leading models for FRB emission. Moreover, for a single burst of FRB 20200120E we  estimated the flux UL from the SuperAGILE detector data in the $18-60$ keV. We performed also a check of the GRID coverage for each burst in the $0.03 - 10$ GeV energy band on short timescales, from $10$ to $10^3$ s, and on longer ones including the complete $\sim$17 years AGILE/GRID archive. We then considered the famous event FRB 200428 from the galactic magnetar SGR 1935+2154 as reference to extrapolate a possible X-ray emission in MCAL and SuperAGILE bands, from the radio energies of R-FRBs using the E$_{\mathrm{X}}$/E$_{\mathrm{radio}}$ of FRB 200428 as fixed parameter. We compared these energies with historical magnetar X-ray bursts rescaled in the same bands. Our observations set useful constraints on the FRB magnetar model in particular, the MCAL ULs are currently the most stringent in the 0.4--30 MeV band.

\end{abstract}

\section{\bf{Introduction}} \label{sec:intro}
\label{intro}

FRBs are intense, millisecond-duration bursts of radio waves originating from extragalactic sources at cosmological distances \citep{Lorimer2007}. Since their discovery in 2007, they have emerged as one of the most enigmatic astrophysical phenomena, with proposed origins ranging from highly magnetized neutron stars (magnetars) to primordial black holes \citep{Popov2018, Zhang2023}. During the past decade, advances in radio telescope technology, particularly from the Canadian Hydrogen Intensity Mapping Experiment (CHIME) telescope and more recently with the Five hundred meter Aperture Spherical Telescope (FAST), have significantly increased the detection rate of FRBs, thereby facilitating a deeper understanding of their population and properties \citep[][]{CHIME2021,Nan}.

A defining characteristic of FRBs is their dispersion measure (DM), which quantifies the total column density of free electrons along the line of sight. DM values usually correlate with distances, as the burst's radio waves are increasingly delayed at lower frequencies while traveling through intergalactic plasma. This property not only allows researchers to estimate the cosmological distance of the sources but also provides a valuable probe of the intervening medium, including the intergalactic baryon distribution \citep{Petroff2019, Macquart2020}.

Among the many mysteries surrounding FRBs, the distinction between two classes of this phenomenon, those detected multiple times, named \textit{repeaters} (R-FRBs) and those detected once, \textit{one-off}, has been a central focus of this field of radio-astronomy. Repeaters, such as FRB 20121102A, FRB 20180916B, and FRB 20220912A, emit multiple bursts over time from the same location, while one-off FRBs appear as single isolated events \citep{Spitler2016, CHIME2023}. Evidence suggests that these two categories may arise from fundamentally different astrophysical processes. Repeaters are often linked to highly magnetized neutron stars (magnetars), whose dynamic magnetospheres or episodic crustal activities could produce recurring emissions \citep{Lyutikov2017}. In contrast, non-repeaters are hypothesized to result from catastrophic events, such as mergers of compact objects or collapse-induced flares \citep{Platts2019, Zhang2023}, or compact objects in binary systems, either X-ray binaries or system including ms-pulsars \citep{Pearlman2025}. Alternately, non-repeaters may be the most intense events of a distribution dominated by weak, and unobservable, events from the same source \citep[][]{Kumar}. Recent discoveries, including the identification of 25 new repeating sources by the CHIME/FRB collaboration, have blurred the distinction between these categories \citep{CHIME2023}. For example, some non-repeaters have been observed to emit additional bursts after long periods of inactivity, raising the possibility that all FRBs might eventually exhibit repetition under certain conditions \citep[][]{Zhang2023}. Furthermore, polarization studies suggest that variations in magnetic field strength and orientation could influence the observed properties of repeaters and non-repeaters, potentially explaining some of their differences \citep{Liu2024}.

Although FRBs have been detected primarily at radio frequencies, the goal to understand their origin and emission mechanisms has led astronomers to conduct extensive multifrequency campaigns. These efforts aim to identify counterparts of FRBs across the electromagnetic spectrum, including optical, X-rays, \textgamma -rays, and even neutrino and GW emissions \citep[][]{Martone, Guidorzi1, Guidorzi2, Piro2021, Principe2023, GW2023, Icecube2023, Ridnaia2024, Pearlman2025}, even with works dedicated to specific source classes \citep[such as short gamma-ray bursts,][]{Curtin2024}. Such campaigns have placed stringent upper limits (ULS) on high-energy counterparts, thereby refining theoretical models of FRB emission mechanisms. Moreover, various further radio campaigns have started later on \citep[][]{Chime2024, Kirsten2024, Ould-Boukattine2024, Konijn2024}, including some from Italian radio telescopes like Northern Cross, Medicina Grueff 32m, Noto 32m, and Sardinia Radio telescope  \citep{Pilia2020, Trudu2022, Pelliciari2024}.

A groundbreaking discovery in this context was the detection of the first Galactic FRB, FRB 200428, associated with the galactic magnetar SGR 1935+2154. This event was observed by multiple instruments, including \agilep, which detected a “burst forest” in the 18--60 keV energy range, followed by a weak X-ray burst coincident with the radio emission of FRB 200428 \citep{Tavani2021b}. This provided the first confirmed high-energy counterpart of an FRB and strong evidence linking magnetars to at least some FRBs \citep{Bochenek2020, Mereghetti2020}.

The \agilep\ satellite, operated by the Italian Space Agency, has played a crucial role in the search for high-energy counterparts of FRBs. Its onboard detectors, \textit{MCAL}, \textit{GRID} and \textit{SuperAGILE}, have been used in systematic searches for \textgamma - and X-ray signals associated with FRBs, yielding important constraints on their eventual high-energy emission \citep{Casentini2020, Tavani2021a, Verrecchia2021}.\\

In this paper, we present a study focused on a selected sample of R-FRBs with DM$_\mathrm{exc} < 300$ pc cm$^{-3}$. This threshold is chosen because, when considering the expected host galaxy contribution to DM and applying the Macquart relation between DM and redshift, the inferred DM$_{\text{igm}}$ values suggest luminosity distances less than about 1 Gpc. Such distances are potentially more favorable for detecting high-energy counterparts with the \agile detectors. This work follows previous ones, the first \citep{Casentini2020} dedicated to two specific R-FRBs (FRB 20180916B and FRB 20181030A) and a second one on a sample of nearby one-off and R-FRBs \citep{Verrecchia2021}. The bursts of the repeating sources included in that works have been re-analyzed for this campaign with updated procedures. For each FRB in this sample, we performed a search for temporal coincidences between known radio bursts and possible \agile hard X-ray and/or \textgamma-ray counterparts. In the following, we will describe the selection of the FRB sample, expose the results of the search inside \agile detectors dataset; a more detailed comparison with magnetar-based models is deferred to the Discussion section, with a detailed comparison of the extrapolated energies with some historical flares from Galactic magnetar flares.

\begin{figure}
   \centerline{
\includegraphics[width=0.70\textwidth, angle = 0]{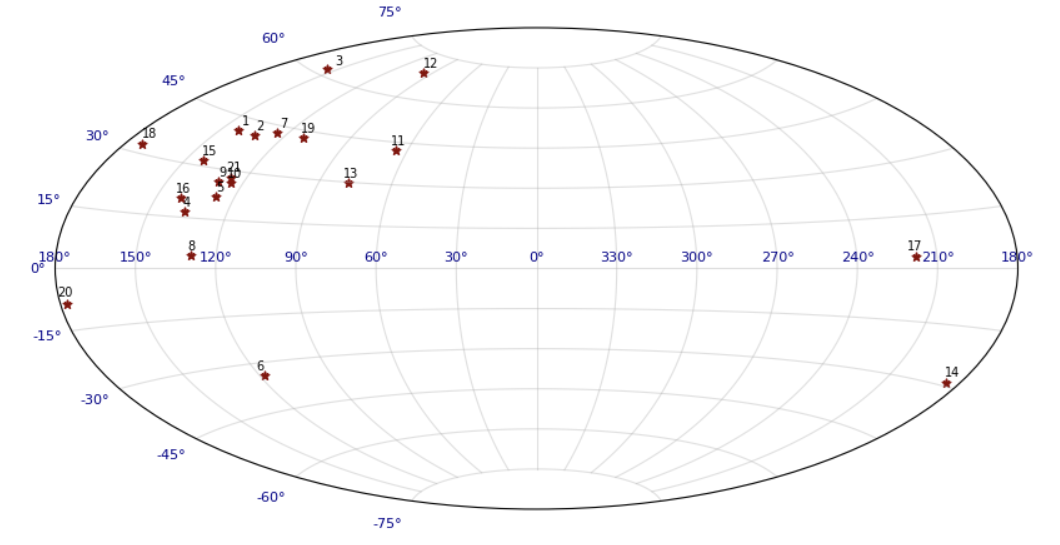}
    }
   \caption{Aitoff projection in Galactic coordinates displaying the sky distribution of the R-FRBs selected in this work. The sources, marked as red stars, exhibit clustering in a specific region of the sky, primarily due to the observational bias introduced by the CHIME radio telescope, which dominates the detection of repeating FRBs. We used numbers to label the sources according their appearance order in Table~\ref{tab:Table_1}.}
 \label{Figure_1}
\end{figure}

\section{\bf{The AGILE satellite}} \label{sec:sec2}
\label{AGILE}

The \agile satellite was an Italian Space Agency mission, launched in 2007 and dedicated to \textgamma-ray astrophysics \citep{Tavani2009}. \agile consisted of four detectors, each sensitive to different energy ranges: hard X-rays ($18 - 60$ keV, \textit{SuperAGILE} imaging detector), MeV energies ($0.35 - 100$ MeV, \textit{MCAL} omnidirectional detector), and GeV \textgamma-ray energies ($0.03 - 50$ GeV, \textit{GRID} imaging detector). An anticoincidence (\textit{AC}) system enclosed the instrument and was capable of detecting hard X-rays in the $50 - 200$\,keV range.
\textit{SuperAGILE} is   
an hard X--ray imager composed of four 1-D coded mask detectors 
with $\sim\,1 sr$ FoV, capable of localizing source with $2-3\arcmin$ accuracy \citep[in pointing mode;][]{Feroci2007}. 
The CsI(Tl) \textit{MCAL} detector \citep[][]{Labanti2009} has a self-triggering capability in "BURST" mode with 7 time windows, allowing the detection of \textgamma-ray transient events at time scales from sub-ms to hundreds of seconds, such as Terrestrial Gamma-ray Flashes (TGF) and GRBs \citep[][]{Marisaldi2014,Maiorana2020,Ursi2022a}.
The \textit{GRID} imager, with a FoV of one-fifth of the sky and $\sim\, 15\arcmin$ location accuracy (for 10\,$\sigma$ detected sources), 
is composed of a Silicon Tracker (ST) capable of reconstructing the source position from the pair conversion of \textgamma-ray photons in a Tungsten layer, the MCAL in "GRID" mode and the AC working together.

Any source within the accessible sky region might have been exposed by the \textit{GRID} for time windows of about $150\,\text{s}$ per revolution, with varying off-axis angles relative to the normal incidence of the \agile instrument. This continuous rotation around its pointing axis enabled coverage of nearly the entire sky every $\sim7$ minutes.

After an extensive and successful mission, the scientific observations of \agile concluded on January 18, 2024. Following the completion of data collection and analysis, the satellite re-entered the Earth's atmosphere on February 13, 2024. This marked the end of over 17 years of contributions to \textgamma-ray astrophysics, during which \agile provided valuable insight into high-energy astrophysical phenomena and significantly advanced our understanding of \textgamma-ray sources.

\section{\bf{Source catalogues and sample selection}} \label{sec:sec3}
\label{sec-3}

The sample of R-FRBs studied in this work was obtained from publicly available online catalogs, which provide comprehensive lists of observed FRBs and their characteristics. Specifically, we used data from the CHIME/FRB catalog\footnote{\url{https://www.chime-frb.ca/}}, the Transient Name Server (TNS)\footnote{\url{https://www.wis-tns.org/}}, and the Blinkverse catalog\footnote{\url{https://blinkverse.alkaidos.cn}}. These repositories store detailed information on each FRB, including DM, coordinates, repetition rates, and other observational properties across multiple instruments. By combining data from these complementary resources, we ensured a more complete and versatile sample for our work. The CHIME/FRB catalog is a key resource for observations in the northern sky, providing data on both repeating and non-repeating sources \citep{CHIME2021}. The TNS serves as a central repository for transient events detected by various observatories and includes a dedicated section for FRBs. Blinkverse \citep{Xu2023} have a collection of more updated data from multiple catalogs (in particular from FAST), making it particularly useful for multiwavelength follow-up studies. Considering these databases, we constructed a sample of nearby R-FRBs, all having small DM$_\mathrm{exc}$ values, as previously reported. To evaluate it, we started from the definition of DM itself, which represents the integrated column density of free electrons along the line of sight. As discussed in \citet{Petroff2019, Cordes2019, Zhang2023}, the observed DM (or \textbf{DM$_\mathrm{tot}$}) of an FRB can be factorized into three main components:
\begin{enumerate}
    \item \textbf{DM$_\mathrm{MW}$}: the contribution from free electrons in the Milky Way, which can be further divided into \textbf{DM$_\mathrm{disk}$} (due to the Galactic disk) and \textbf{DM$_\mathrm{halo}$}. Galactic disk models are provided by studies such as \citet{NE2001, YMW16}, while for the halo a typical value of $30 \, \mathrm{pc \, cm^{-3}}$ is commonly assumed.
    \item \textbf{DM$_\mathrm{IGM}$}: the contribution from free electrons in the intergalactic medium (IGM).
    \item \textbf{DM$_\mathrm{host}$}: the contribution from the host galaxy, including any local environment near the source. Due to the difficulty in pinpointing and resolving the host galaxy, this component is often approximated as $50\, (1+z) \, \mathrm{pc \, cm^{-3}}$ \citep{Petroff2019, Cordes2019}.
\end{enumerate}
\begin{deluxetable*}{c|c|c|c|c|c|c}
\tablecaption{Properties of the R-FRBs considered in this work. The first column lists the names of the FRB sources. The second column provides $\mathrm{DM_{exc}}$, with the mean $\mathrm{DM_{tot}}$ given in parentheses. The third column shows the Galactic coordinates of the sources in degrees. The fourth column $N_b$ indicates the total number of bursts observed to date for each source. The fifth and sixth columns give for each source the minimum and maximum burst widths in milliseconds and the minimum and maximum radio fluence values in $\mathrm{Jy \, ms}$, respectively. The seventh column lists the spectroscopic redshift $z_s$ associated with each source (where available), with the estimated luminosity distance ($D_L$) in Mpc in parentheses.
\label{tab:Table_1}}
\tabletypesize{\small} 
\tablewidth{0.9\textwidth} 
\tablehead{ 
\colhead{\textbf{FRB}} & 
\colhead{\textbf{DM$_{\rm exc}$}} & 
\colhead{\textbf{\itshape{l}, \itshape{b}}} & 
\colhead{\textbf{N$_{\rm b}$}} & 
\colhead{\textbf{Width}} & 
\colhead{\textbf{{\itshape f}}} & 
\colhead{\textbf{z$_{\rm s}$}}\\
\colhead{\textbf{Name}} &
\colhead{\textbf{(pc$\cdot$cm$^{-3}$)}} & 
\colhead{\textbf{(deg)}} & 
\colhead{\textbf{}} & 
\colhead{\textbf{(ms)}} & 
\colhead{\textbf{(Jy ms)}} & 
\colhead{\textbf{- (Mpc)}}
}
\startdata
FRB 20200120E & 18.0 (88.0)    & $142.19^\circ, +41.22^\circ$  & 80  & $0.01 - 0.70$  & $0.04 - 2.40$  & $0.0008\, (3.6)$ \\
FRB 20181030A & 33.5 (103.7)   & $133.40^\circ, +40.92^\circ$  & 9  & $0.59 - 1.43$  & $4.50 - 7.30$  & $0.004\, (17.3)$  \\
FRB 20220912A & 67.1 (222.1)   & $155.41^\circ, +61.81^\circ$  & 1529  & $0.09 - 32.9$  & $0.002 - 37.7$  & $0.077\, (343)$  \\
FRB 20180814A & 72.4 (189.4)   & $136.42^\circ, +16.60^\circ$  & 22  & $7.90 - 63.0$  & $3.40 - 66.0$  & $0.078\, (347)$  \\
FRB 20190107B & 73.1 (173.1)   & $127.14^\circ, +21.75^\circ$  & 2  & $0.98$  & $4.30$  & $0.11\, (500)$    \\
FRB 20200223B & 125.5 (201.5)   & $118.07^\circ, -33.88^\circ$  & 10  & $1.26 - 2.29$  & $1.06 - 14.4$  & $-$    \\
FRB 20180908B & 127.8 (195.8)   & $124.74^\circ, +42.86^\circ$  & 4  & $1.60 - 9.00$  & $1.10 - 2.90$  & $0.17\, (800)$    \\
FRB 20180916B & 150.7 (349.4)   & $129.71^\circ, +3.74^\circ$   & 290  & $0.06 - 158$  & $0.08 - 318$  & $0.033\,(150)$   \\
FRB 20181226F & 152.1 (240.1)   & $129.79^\circ, +26.20^\circ$  & 3  & $-$  & $-$  & $-$    \\
FRB 20190905A & 153.3 (240.4)   & $124.64^\circ, +26.25^\circ$  & 6  & $0.75 - 1.13$  & $1.79 - 18.0$  & $-$    \\
FRB 20190110C & 156.3 (222.0)   &  $65.56^\circ, +42.15^\circ$   & 3  & $2.95$  & $1.40$  & $0.22\, (1050)$    \\
FRB 20190303A & 164.1 (223.1)   &  $97.48^\circ, +68.94^\circ$   & 38  & $0.89 - 7.20$      & $0.78 - 10.4$  & $0.064\, (285)$   \\
FRB 20190812A & 170.7 (249.1)   &  $78.05^\circ, +29.83^\circ$   & 2  & $0.42 - 0.60$       & $0.65 - 13.0$  & $-$    \\
FRB 20201130A & 202.3 (288.3)   & $185.33^\circ, -29.15^\circ$  & 12  & $0.47 - 6.55$      & $1.31 - 20.4$  & $-$    \\
FRB 20210323C & 204.7 (285.2)   & $142.59^\circ, +31.54^\circ$  & 11  & $0.83 - 5.00$      & $2.32 - 7.60$  & $-$    \\
FRB 20191105B & 210.3 (313.8)   & $140.59^\circ, +20.40^\circ$  & 2  & $0.55 - 0.71$       & $2.70 - 19.7$  & $-$    \\
FRB 20190113A & 217.6 (426.5)   & $218.04^\circ, +3.43^\circ$   & 3  & $1.82 - 3.03$       & $5.70 - 9.40$  & $0.29\, (1450)$    \\
FRB 20190907A & 226.8 (309.8)   & $173.39^\circ, +32.26^\circ$  & 7  & $0.54 - 3.00$       & $0.70 - 6.90$  & $-$    \\
FRB 20201114A & 253.9 (321.9)   & $111.18^\circ, +42.63^\circ$  & 2  & $0.72 - 1.04$       & $2.45 - 4.70$  & $-$    \\
FRB 20201124A & 257.6 (414.1)   & $177.60^\circ, -8.50^\circ$   & 2883  & $0.91 - 316$    & $0.005 - 640$  & $0.098\, (445)$   \\
FRB 20200127B & 267.0 (351.0)   & $126.63^\circ, +28.02^\circ$  & 2  & $0.45 - 0.61$       & $4.70 - 7.70$  & $-$    \\ \enddata
\end{deluxetable*}
The intergalactic component, $\mathrm{DM_{IGM}} = \mathrm{DM_{tot}} - \mathrm{DM_{MW}} - \mathrm{DM_{host}}$, is particularly important because it can be related to the redshift $z$ of the FRB through the Macquart relation, expressed as:
\begin{equation}
\mathrm{DM_{IGM}} \approx 900 \, z \; \mathrm{pc \, cm^{-3}}.
\label{eq:macquart_relation}
\end{equation}

This relation allows to estimate cosmological distances and use FRBs as probes of baryonic matter in the IGM \citep{Macquart2020}. Given the challenges in separating DM$_\mathrm{host}$ from DM$_\mathrm{IGM}$, the combined parameter DM$_\mathrm{exc}$, defined as DM$_\mathrm{IGM} +$ DM$_\mathrm{host}$, is commonly adopted in literature. DM$_\mathrm{exc}$ serves as a useful proxy for distance, especially for sources without well-constrained host associations.

\begin{figure}
   \centerline{
        \includegraphics[width=0.70\textwidth, angle = 0]{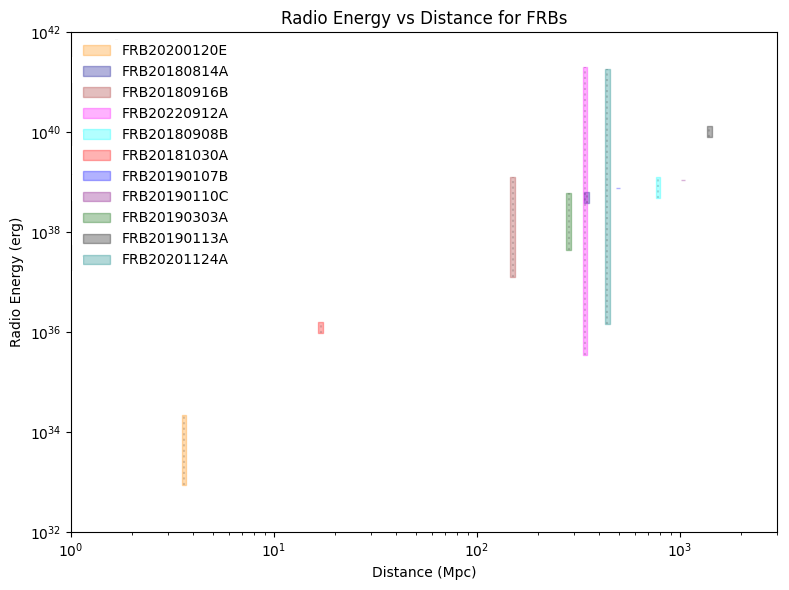}}

   \caption{Radio energy range of the R-FRBs selected in this work, derived using the fluence values from Table~\ref{tab:Table_1}. For each repeater, only the minimum and maximum fluence values were considered to estimate the range of energies at which each source has been observed to date. To better highlight individual repeaters, stripes of arbitrarily increased size have been added at their distance, creating distinct colored boxes.}
   \label{Figure_2}
\end{figure}

As previously mentioned, here we selected R-FRBs with DM$_\mathrm{exc} < 300 \, \mathrm{pc \, cm^{-3}}$. The selected sources are listed in Table~\ref{tab:Table_1} and shown in Galactic coordinates in Figure~\ref{Figure_1}. This threshold was chosen to emphasize relatively nearby sources, whose lower DM$_\mathrm{exc}$ values reduce uncertainty in the inferred distance and increase the likelihood that their high-energy counterparts, if present, could be detected by MCAL, GRID and SuperAGILE instruments. More distant FRBs, with larger DM$_\mathrm{exc}$ values, would likely be too faint for detection in the energy ranges covered by \agile. We have also evaluated the radio energy range per each R-FRB that has known distance. It was evaluated using equation

\begin{equation}
    E_{\mathrm{radio}} \ [\mathrm{erg}] = 4\pi f \, \nu \, \mathrm{d^2_{FRB}}
    \label{eq:radio_energy}
\end{equation}

in the main assumption of isotropic emission. In Equation~\eqref{eq:radio_energy}, $f$ is the fluences in Table~\ref{tab:Table_1} (converted to $\text{erg}\,\text{cm}^{-2}\,\text{Hz}^{-1}$), $\nu$ is the center frequency of the observational bandwidth (e.g., $600 \, \mathrm{MHz}$ for CHIME's $400-800 \, \mathrm{MHz}$ bandwidth), and $\mathrm{d_{FRB}}$ is the luminosity distance derived from the redshift. We considered, per each source, only the maximum and minimum fluences to obtain a range in energies. The results are reported in Figure~\ref{Figure_2}. An obvious distance-energy relation appears in this plot, with closer sources showing lower energies with respect to the more distant ones. Actually, this is only an observational bias due to the sensitivity of the radio telescopes. On the other hand, we observe the lack of very luminous radio bursts at closer distances. This could be explained in several way: geometry of the source, rarity of the superluminous events or, according to some authors \citep{Zhang2023}, different ages of the emitters.

\section{\bf{MCAL Observations}} \label{sec:sec4}
\label{sec-4}

\begin{figure}
   \centerline{
        \includegraphics[width=0.70\textwidth, angle = 0]{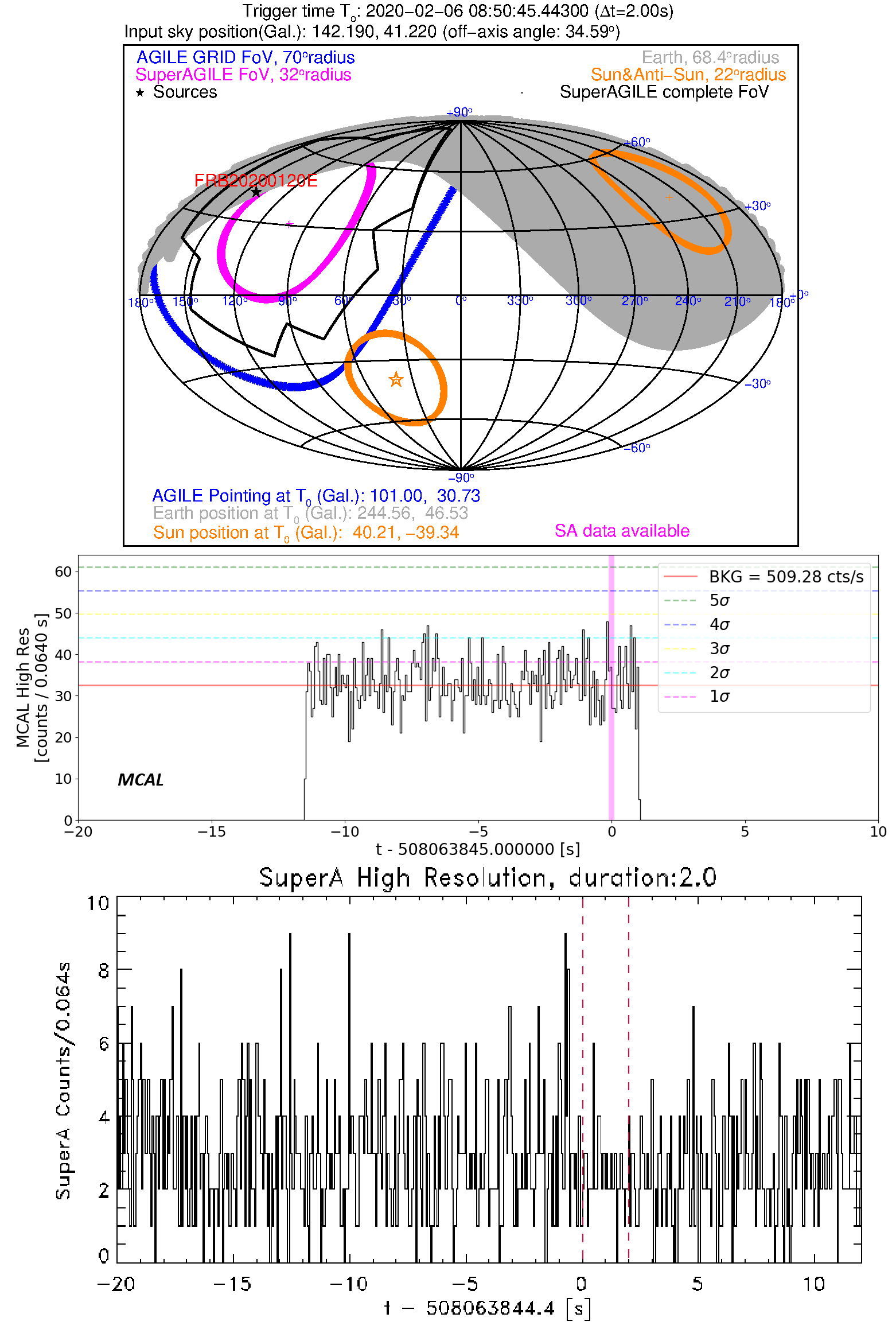}}
   \caption{\textit{Top panel:} FoV of the \agile satellite, illustrating the coverage of its detectors for one burst of FRB 20200120E. The shaded gray region represents the Earth’s occultation, while the orange circle indicates the position of the Sun/AntiSun region. The SuperAGILE fully coded FoV part is shown with a magenta circle while the complete FoV has a cross shape in black; the GRID FoV is a blue circle, whereas the coverage of MCAL is constrained only by Earth’s occultation. The black star marks the position of the source. \textit{Middle panel:} MCAL light curve of the event in the $0.4-30$ MeV. \textit{Bottom panel:} SuperAGILE lightcurve of the event in the $18-60$ keV. Both light curves have a 64 ms binning. Vertical dashed lines indicates the time interval used to evaluate the UL.}
   \label{Figure_3}
\end{figure}

We considered, for each FRB in the sample, only those bursts for which the topocentric arrival time (at infinite or finite frequency) was available. This requirement excluded several known bursts, particularly those observed by the FAST telescope, reducing the repetition sample from $4920$ bursts to $782$. The second step in our analysis was to verify whether each burst occurred when the source sky-position fell inside the FoV of the \agile detectors. We considered an event to be visible for MCAL detector if its position was not occulted by Earth at each burst time. To determine the MCAL visibility region, in this work, we considered a circular Earth occultation region with radius of $68.4^\circ$. An example of FoV is shown in Figure~\ref{Figure_3} for one of the burst of the source FRB 20200120E, where the black star represents the source location.
\begin{deluxetable*}{c|c|c|c|c}
    \tablecaption{Summary of the selected bursts, their visibility in MCAL and the evaluated ULs. The first column lists the R-FRB, the second column provides the number of repetitions for each R-FRB selected after the exclusion of those not having all the radio informations required. The third column gives the number of events in visibility for MCAL, with the percentage shown in parentheses. The fourth column indicates the number of bursts for which MCAL was in visibility and simultaneous high-resolution data from the calorimeter was available. The fifth column lists the ULs in fluence for all bursts not having high-resolution data, based on the onboard triggering logic thresholds.}

    \label{tab:Table_2}
    \tablewidth{0pt}
    \tablehead{
    \colhead{\textbf{FRB}} & 
    \colhead{\textbf{Selected}} & 
    \colhead{\textbf{Within}} & 
    \colhead{\textbf{Data}} &
    \colhead{\textbf{MCAL}}\\
    \colhead{\textbf{Name}} & 
    \colhead{\textbf{bursts}} & 
    \colhead{\textbf{MCAL FoV}} &
    \colhead{\textbf{@T$_{\rm burst}$}} &
    \colhead{\textbf{$\mathscr{F}$ UL}}\\
    \colhead{} & 
    \colhead{} & 
    \colhead{} & 
    \colhead{} &
    \colhead{\textbf{$\times 10^{-8} \rm [erg / cm^{2}]$}}
    }
    \startdata
        FRB 20200120E & $13$  & $10$ ($77\%$) & $1$ ($\sim 8\%$) & $-$\\
        FRB 20181030A & $9$  & $9$ ($100\%$) & $-$ & $1.7-9.2$\\
        FRB 20220912A & $469$  & $311$ ($66\%$) & $15$ ($\sim 3\%$) & $-$\\
        FRB 20180814A & $22$  & $19$ ($86\%$) & $2$ ($\sim 9\%$) & $-$\\
        FRB 20190107B & $2$  & $2$ ($100\%$) & $-$ & $3.1-5.2$\\
        FRB 20200223B & $9$ & $4$ ($44\%$) & $-$ & $1.8-4.2$\\
        FRB 20180908B & $4$  & $4$ ($100\%$) & $1$ ($\sim 25\%$) & $-$\\
        FRB 20180916B & $136$  & $108$ ($79\%$) & $6$ ($\sim 4\%$) & $-$\\
        FRB 20181226F & $3$  & $3$ ($100\%$) & $-$ & $3.0-4.3$\\
        FRB 20190905A & $6$  & $6$ ($100\%$) & $-$ & $1.6-3.2$\\
        FRB 20190110C & $3$  & $3$ ($100\%$) & $-$ & $1.8-2.6$\\
        FRB 20190303A & $34$  & $21$ ($62\%$) & $-$ & $1.8-9.3$\\
        FRB 20190812A & $2$  & $2$ ($100\%$) & $-$ & $1.8-1.9$\\
        FRB 20201130A & $12$  & $9$ ($75\%$) & $-$ & $1.8-5.3$\\
        FRB 20210323C & $10$  & $5$ ($50\%$) & $-$ & $3.0-4.2$\\
        FRB 20191105B & $2$  & $2$ ($100\%$) & $-$ & $2.9-3.1$\\
        FRB 20190113A & $3$  & $1$ ($33\%$) & $-$ & $3.2-5.1$\\
        FRB 20190907A & $7$  & $2$ ($29\%$) & $-$ & $1.6-5.2$\\
        FRB 20201114A & $2$  & $2$ ($100\%$) & $-$ & $1.5-1.8$\\
        FRB 20201124A & $34$  & $14$ ($41\%$) & $-$ & $1.7-5.3$\\
        FRB 20200127B & $2$  & $2$ ($100\%$) & $-$ & $1.9-4.3$\\
    \enddata
\end{deluxetable*}

This initial visibility check significantly reduced the repetition sample available for each R-FRB. The results of this step, reported in third column of Table~\ref{tab:Table_2}, reduce the sample of bursts to $529$ repetitions to be considered. Once we selected the exposed bursts, we proceeded to determine if there is triggered and acquired MCAL high-resolution data at each repetition time.

MCAL triggering system acquired high resolution data ("BURST" mode) by applying seven different trigger search integration times (SITs), in three energy bands. The data are managed by seven trigger logics, ranging from $0.293$ up to $8192$ ms. Data acquisition starts whenever a count rate among these logics overcomes the set thresholds \citep[for more details see][]{Casentini2020,Ursi2022}. 

For the majority of the repetitions within the MCAL FoV ($\sim97\%$) we found no coincident data acquisitions at FRB burst times (within T$_{\mathrm{burst}}\pm\,1\,$s interval), while for the remaining $3\%$ we have temporally coincident data: specifically, we identified 1 coincident trigger for FRB 20200120E, 15 for FRB 20220912A, 2 for FRB 20180814A, 1 for FRB 20180908B, and 6 for FRB 20180916B. A summary of these bursts is reported in Table~\ref{tab:Table_3}. An example of a high resolution data chunk ($\sim 10$ s) including a burst time, is shown in the middle panel of Figure~\ref{Figure_3} for the only FRB 20200120E, burst covered by MCAL. We conducted an individual analysis of each trigger but we did not found evidence of a true source while these features appear to be consistent with background fluctuations. Moreover, as verification, we calculate the FAR of the MCAL high-resolution data acquisition intervals, analysing a 100-day period before each involved trigger. The FAR was approximately $1.2 \times 10^{-3}$\,Hz, which is consistent with MCAL acquisition due to background fluctuations.

Since we found no genuine astrophysical signal in the data at the FRB burst times, we estimated flux and fluence upper limits (ULs) for potential hard X-ray counterparts. For bursts without data acquisition, we focused on the 1 ms trigger logic, appropriate for the characteristic timescales of FRB. A signal would require at least 8 counts to exceed the threshold in this logic. From this requirement, we inferred the minimum fluence needed to produce a trigger at the known burst time and source position, taking into account the \agile orientation and appropriate response matrices. The resulting ULs are presented in the fifth column of Table~\ref{tab:Table_2}.

We estimated the flux needed for a $3\sigma$ detection for the 25 bursts finally selected, in the hypothesis of a cutoff power-law (CPL) spectral model, following the approach described in \citet{Pelliciari2024}. This estimation is shown in the fifth column of Table~\ref{tab:Table_3}.  While a detailed comparison with magnetar-based emission models will be discussed in Section~\ref{sec:sec7}, our preliminary tests suggest that the derived flux ULs are relatively insensitive to model parameters variation and provide robust constraints on the non-detection of high-energy counterparts to these FRBs in the MCAL band.

\begin{figure}
 \centerline{
 \hspace{1.3cm}
        \includegraphics[width=0.60\textwidth, angle = 0]{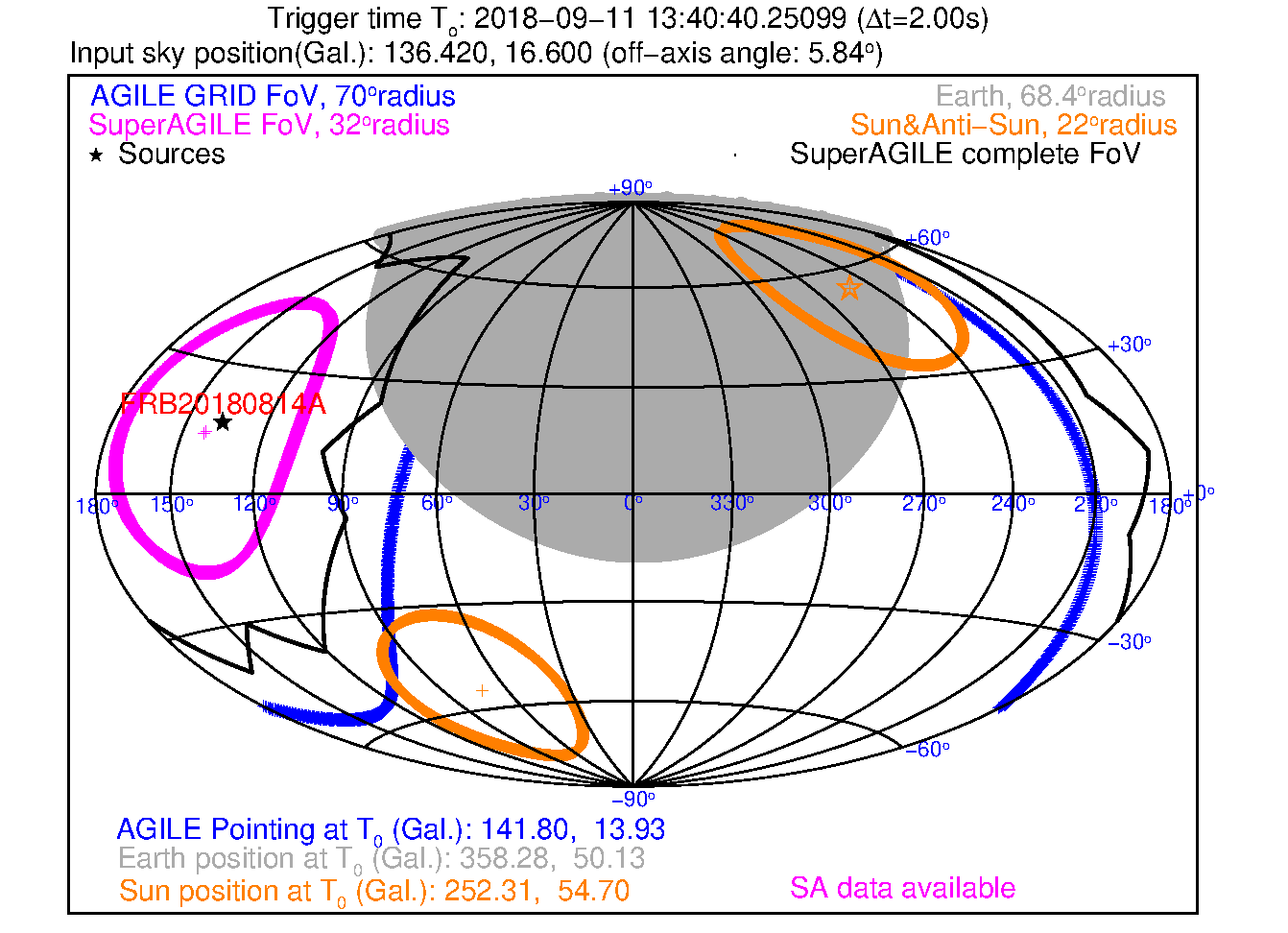}}
\centerline{
        \includegraphics[width=0.60\textwidth, angle = 0]{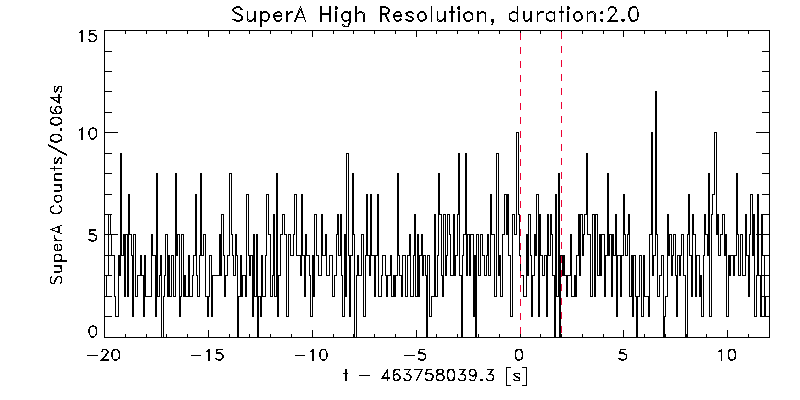}}
   \caption{The FoV of the \agile satellite showing the coverage of its detectors for one burst of FRB 20180814A (upper panel), and the corresponding SuperAGILE light curve in the $18-60$ keV (bottom panel), respectively. The light curve have a $64$ ms binning and vertical dashed lines indicates the time interval used to evaluate the UL.}
   \label{Figure_4}
\end{figure}

\section{\bf{GRID and SuperAGILE observations}} \label{sec:sec5}
\label{GRID}
We checked the AGILE/GRID exposure coverage of the bursts, verifying if the source fell within the FoV and, as for MCAL, excluding Earth-occulted regions (see for instance Figure~\ref{Figure_3}), and also time intervals corresponding to SAA (South Atlantic Anomaly) passages. We found 131 burst within the GRID FoV (with radius 70$\degree$) at each T$_0$.
We applied the two usual procedures to estimate possible detection and/or to extract Flux ULs \citep{Verrecchia2021, Ursi2022}, the "GRB detection mode" for short time-scales \citep[1-1000 s;][]{2010ApJ...708L..84G, Verrecchia2021}, and the AGILE Maximum Likelihood \citep[AML;][]{2012Bulgarelli} for the long ones ($\rm days - \rm years$). We selected reliable events at angles $> 80\,\deg$ from Earth center and at off-axis angle $<\,60\,\deg$.
The short timescales analysis, including also events having lower probabilities to be photons, allowed us to obtain a few low pre-trial significance detections applying the detection algorithm in \citet{1983Li&Ma}, which we discarded as not reliable due to their position at the border of the exposed FoV, and taking into account the post-trial correction. 
We evaluated the ULs in "GRB mode" in the 50 MeV – 10 GeV energy band with the creation of the short time scale $\gamma$-ray sensitivity maps \citep[][]{Tavani2016,Verrecchia2017}, obtaining the 2\,$\sigma$ flux UL values reported in Table~\ref{tab:Table_5}, ranging from $10^{-9}\,\rm erg\,cm^{-2}\,s^{-1}$ up to $10^{-6}\,\rm erg\,cm^{-2}\,s^{-1}$ about for $1000$ and $10$\,s exposures.
Moreover, for the $10$\,s integration regarding FRB 20200120E, we performed a spectral modeling with the Xspec software \citep[][within the HEAsoft package v.6.26.1]{Xspec}, considering other GRID detections of GRBs as reference\footnote{for instance GRB 080514B \citep{2008A&A...491L..25G}.}. We applied a power law spectral model with photon indices from $-2.0$ to $-2.5$ and $-3.0$ obtaining the upper limits $1.1\times10^{-7}\,\mathrm{erg\,cm^{-2}\,s^{-1}}$, $1.8\times10^{-7}\,\mathrm{erg\,cm^{-2}\,s^{-1}}$ and $2.0\times10^{-7}\,\mathrm{erg\,cm^{-2}\,s^{-1}}$, respectively; see Figure~\ref{Figure_7}.

In the long time scale case, we analysed integrations of $\sim\,$5 and finally $\sim\,$17 years. We did not find any reliable detection, after discarding emission from some nearby contaminating source not included in the Second AGILE/GRID source catalogue \citep{2019Bulgarelli}, and 2\,$\sigma$ flux ULs were extracted in the 100 MeV – 10 GeV band applying the AML analysis at the source position considering a standard power-law spectral model, with a mean photon index of -2.0. The AML is based on the likelihood ratio test, as the ratio of the ML of the source presence hypothesis versus the ML of the null hypothesis, both build from the counts, exposure and Galactic diffuse emission maps within 10 degrees, and taking into account the detector response \citep{2012Bulgarelli,2019Bulgarelli}. In our case, in the GRID band we do not have indication of an expected spectral model, so we decided to test also two softer spectral index values, -2.5 and -3.0. We obtained, with the standard index value, 2\,$\sigma$ flux ULs ranging from $2.9\,\times\,10^{-13}\,\rm erg\,cm^{-2}\,s^{-1}$ up to $2.9\,\times\,10^{-12}\,\rm erg\,cm^{-2}\,s^{-1}$ on 17 years integrations. Applying the softer values -2.5 and -3.0, we obtained values a factor of 1.26 and 1.67 higher for the minimum UL, and a factor of 2 and 3 higher for the maximum one, respectively. We can conclude that for softer values the order of magnitude remains the same.

We checked also the SuperAGILE imager coverage of the bursts selected, within the "fully-coded" FoV part only, finding six covered bursts, and moreover a single burst of FRB 20200120E\, falling just outside the fully coded FoV border. The complete list of covered bursts and their related flux UL in the SuperAGILE band are reported in Table~\ref{tab:Table_4}.

We analysed the SuperAGILE high resolution data available for these events within $\pm\,100 \rm s$, after the exclusion of time intervals including SAA passages, and generated light curves in three binning: 1, 0.5 and 0.064 s, with the "burst" pipeline \citep{Feroci2007}. We did not find any significant, reliable detection for the six bursts and we estimated 3\,$\sigma$ flux ULs ranging from $4.4\times 10^{-8}\,\rm  erg\,cm^{-2}\,s^{-1}$ to $9.0\,\times \,10^{-8}\,\rm  erg\,cm^{-2}\,s^{-1}$, where the minimum value is obtained for the FRB 20180814A burst falling at an off-axis angle of only $\sim\,6\degree$.

The light curve with a $0.064$ s binning for the FRB 20200120E burst is shown in the bottom panel of Figure~\ref{Figure_3}. The source was at about 35$^{\circ}$ off-axis. We did not find any detection but we could estimate a 3\,$\sigma$ flux UL of $9.0\, \times 10^{-8}\,\rm  erg\,cm^{-2}\,s^{-1}$ in the 18\,--\,60 keV band. Moreover we show the \agile FoV and the $0.064$ s binned light curve of the FRB 20180814A burst in Figure~\ref{Figure_4}.\\

\begin{deluxetable*}{c|c|c|c|c}
\tablecaption{List of 25 bursts of R-FRBs in temporal coincidence with MCAL data acquisition. In column five we report the related $3$\textsigma\, flux ULs in [0.4-30] MeV energy band.}
\label{tab:Table_3}
\tabletypesize{\small}
\tablewidth{0.9\textwidth}
\tablehead{ 
\colhead{\textbf{FRB}} & 
\colhead{\textbf{Burst}} &
\colhead{\textbf{Telescope}} & 
\colhead{\textbf{Topocentric MJD}} &
\colhead{\textbf{F$_{[0.4-30]\mathrm{MeV}}$ UL}}\\
\colhead{\textbf{Name}} &
\colhead{\textbf{ID}} &
\colhead{\textbf{Name}} &
\colhead{\textbf{@ infinite freq.}} &
\colhead{\textbf{$\times 10^{-8} \rm [erg / cm^{2} / s]$}}
}
\startdata
FRB 20200120E & $200206$ & CHIME & $58885.36858507$ & $2.57$\\
\hline
FRB 20220912A & $221020$ & CHIME & $59872.22447255$ & $1.28$\\
FRB 20220912A & $221023$D & CHIME & $59875.20976605$ & $3.00$\\
FRB 20220912A & $221107$E & CHIME & $59890.17072415$ & $3.60$\\
FRB 20220912A & $221114$C & CHIME & $59897.15290155$ & $3.60$\\
FRB 20220912A & $221203$H & CHIME & $59916.09782005$ & $1.50$\\
FRB 20220912A & $221220$D & CHIME & $59933.05681718$ & $3.60$\\
FRB 20220912A & $230201$ & CHIME & $59976.93469302$ & $8.99$\\
FRB 20220912A & $230407$B & CHIME & $60041.75513516$ & $2.57$\\
FRB 20220912A & $230414$ & CHIME & $60048.73058087$ & $5.99$\\
FRB 20220912A & $230418$ & CHIME & $60052.72499415$ & $4.49$\\
FRB 20220912A & $230504$B & CHIME & $60068.67979849$ & $2.25$\\
FRB 20220912A & $230510$B & CHIME & $60074.66564847$ & $3.00$\\
FRB 20220912A & B$09$ & Croce del Nord & $60166.04590747$ & $2.06$\\
FRB 20220912A & B$10$ & Croce del Nord & $60166.04590824$ & $2.04$\\
FRB 20220912A & B$11$ & Croce del Nord & $60173.03121566$ & $2.00$\\
\hline
FRB 20220814A & $180911$ & CHIME & $58372.56991025$ & $1.63$\\
FRB 20220814A & $180919$ & CHIME & $58380.52513716$ & $2.57$\\
\hline
FRB 20180908B & $190718$ & CHIME & $58682.04953763$ & $3.00$\\
\hline
FRB 20180916B & $210129$A & CHIME & $59243.04936674$ & $2.00$\\
FRB 20180916B & SRT-P-$11$ & SRT & $59244.93891227$ & $1.07$\\
FRB 20180916B & uGMRT-$01$ & uGMRT & $59441.02524327$ & $1.03$\\
FRB 20180916B & uGMRT-$06$ & uGMRT & $59441.09399652$ & $1.09$\\
FRB 20180916B & uGMRT-$07$ & uGMRT & $59441.10061893$ & $2.36$\\
FRB 20180916B & $221231$ & CHIME & $59944.14644154$ & $1.50$\\
\enddata
\end{deluxetable*}

\begin{deluxetable*}{c|c|c}
\tablecaption{R-FRBs GRID short-time scale $2\,\sigma$ Flux ULs in the 50 MeV\,–\,10 GeV energy band, when available. Start/Stop times are in seconds with respect to each burst time.}
\label{tab:Table_5}
\tablehead{
\colhead{\textbf{FRB}} & 
\colhead{\textbf{pre-T0 min/max value}} & 
\colhead{\textbf{post-T0 min/max value}} \\
\colhead{\textbf{Name}} &
\colhead{Tstart: $-1000$ s; $-100$ s; $-10$ s} & 
\colhead{Tstop: $10$ s; $100$ s; $1000$ s} \\
\colhead{ } &
\multicolumn{2}{c}{\textbf{$\times 10^{-7}\,\mathrm{[erg/cm^{2}/s]}$}}
}
\startdata
FRB 20200120E & $0.085-0.320$ ; $0.049-0.850$ ; $-$ & $1.400$ ;  $0.410-1.800$ ; $0.083-1.400$  \\
FRB 20181030A & $0.069-0.210$ ; $0.250-3.500$ ; $2.300$ &  $5.100$ ;  $2.600-12.000$ ; $0.070-1.100$  \\
FRB 20220912A & $0.061-4.331$ ; $0.221-7.019$ ; $1.168-7.019$ & $1.203-7.006$ ; $0.202-8.748$ ; $0.064-7.032$ \\
FRB 20180814A & $0.063-0.174$ ; $0.299-14.076$ ; $-$  & $1.194-7.036$ ; $0.236-1.418$ ; $0.064-0.201$ \\
FRB 20190107B & $0.086-0.307$ ; $0.479-0.768$ ; $1.456-1.456$ & $1.272-1.284$ ; $0.217-0.255$ ; $0.058-0.079$  \\
FRB 20200223B & $0.247$ ;  $-$ ; $-$ & $-$ ; $-$ ; $-$  \\
FRB 20180908B & $0.072-4.333$ ; $0.271-4.500$ ; $1.447$ & $2.142$ ; $0.437-1.896$ ; $0.078-0.224$  \\
FRB 20180916B & $0.061-4.548$ ; $0.193-9.379$ ; $0.902-7.064$ & $0.902-7.062$ ; $0.233-3.531$ ; $0.067-7.005$  \\
FRB 20181226F & $0.088-0.202$ ; $0.402$ ; $-$ & $1.291$ ; $0.340-0.698$ ; $0.072-0.080$  \\
FRB 20190905A & $0.061-0.083$ ; $0.244-3.345$ ; $3.764$ & $1.344-1.733$ ; $0.226-0.526$; $0.058-0.091$ \\
FRB 20190110C & $0.083-0.181$ ;  $-$ ; $-$ & $-$  ; $0.530$ ; $0.071$  \\
FRB 20190303A & $0.078-6.904$ ; $1.063-7.047$ ; $4.205-7.047$ & $1.400-2.348$ ; $0.214-1.173$ ; $0.060-13.264$ \\
FRB 20190812A & $0.101$ ; $0.232$ ; $-$ & $2.391$ ; $1.196$ ; $0.084$ \\
FRB 20201130A & $0.073-0.216$ ;  $0.323$ ; $7.031$ & $-$  ;  $-$ ; $0.145$  \\
FRB 20210323C & $0.082-0.168$ ; $0.168$ ; $0.168$ &  $0.168$ ; $0.168$; $0.065-0.168$  \\
FRB 20191105B & $0.079$ ; $0.338$ ; $1.303$ & $1.305$ ; $0.460$ ; $0.221$ \\
FRB 20190113A & $-$ ; $-$ ; $-$ & $-$ ; $-$ ; $-$  \\
FRB 20190907A & $0.105$ ; $7.081$ ; $-$ & $4.939$ ; $0.279$ ; $0.235$  \\
FRB 20201114A & $-$ ;  $-$ ; $-$ & $-$ ; $-$ ; $-$  \\
FRB 20201124A & $0.059-1.615$ ; $0.206-0.498$ ; $0.498-1.863$ & $0.498-4.153$ ; $0.498-2.076$; $0.081-0.498$ \\
FRB 20200127B & $0.073-0.092$ ; $0.311$ ; $-$ & $-$ ; $1.189$ ; $0.078-0.091$
1\enddata
\end{deluxetable*}

\begin{deluxetable*}{c|c|c|c|c}
\tablecaption{List of the bursts of R-FRBs in the SuperAGILE FoV and covered by its real-time data acquisition. We include also the case of the burst of FRB20200120E (marked here with *), showed in Figure~\ref{Figure_3}, even if the event occurs when the Source was on the edge of the SuperAGILE fully-coded mask. In fifth column we report the related $3$\textsigma\, flux ULs in [18-60] keV energy band.}
\label{tab:Table_4}
\tabletypesize{\small}
\tablewidth{0.9\textwidth}
\tablehead{ 
\colhead{\textbf{FRB}} & 
\colhead{\textbf{Burst}} &
\colhead{\textbf{Telescope}} & 
\colhead{\textbf{Topocentric MJD}} &
\colhead{\textbf{F$_{[18-60]\mathrm{keV}}$ UL}}\\
\colhead{\textbf{Name}} &
\colhead{\textbf{ID}} &
\colhead{\textbf{Name}} &
\colhead{\textbf{@ infinite freq.}} &
\colhead{\textbf{$\times 10^{-8} \rm [erg / cm^{2} / s]$}}
}
\startdata
FRB 20200120E (*) & $200206$ & CHIME & $58885.36858507$ & 9.0\\
\hline
FRB 20180814A & $180911$ & CHIME & $58372.56991025$ & 4.4\\
\hline
FRB 20190107B & $190107$ & CHIME & $58490.14044364$ & 5.2\\
FRB 20190107B & $190308$ & CHIME & $58550.46699474$ & 6.6\\
\hline
FRB 20180916B & $220328$ & CHIME & $59666.89952457$ & 5.8\\
\hline
FRB 20181226F & $190201$ & CHIME & $58515.65015015$ & 5.7\\
\hline
FRB 20191105B & $191105$ & CHIME & $58792.44450878$ & 5.3
\enddata
\end{deluxetable*}

\section{\bf{Discussion}} \label{sec:sec7}
\label{discussion}

The radio emission from FRBs is thought to represent only a small fraction of the total energy associated with an isotropically emitting plasma configuration. For high-DM FRBs, the estimated energy in the radio band can be of the order of $10^{39-40}$ erg or higher \citep{Cordes2019}. If these high-DM FRBs are extragalactic, their energy sources are likely linked to compact objects, such as neutron stars and/or black holes. On the other hand low-DM FRBs provide a unique opportunity to study energy and luminosity on scales significantly lower than those inferred for extragalactic sources. Assuming isotropic emission, we derive in the following observational constraints on the energetics.

\begin{figure}[h]
   \centerline{
\includegraphics[width=0.7\textwidth, angle = 0]{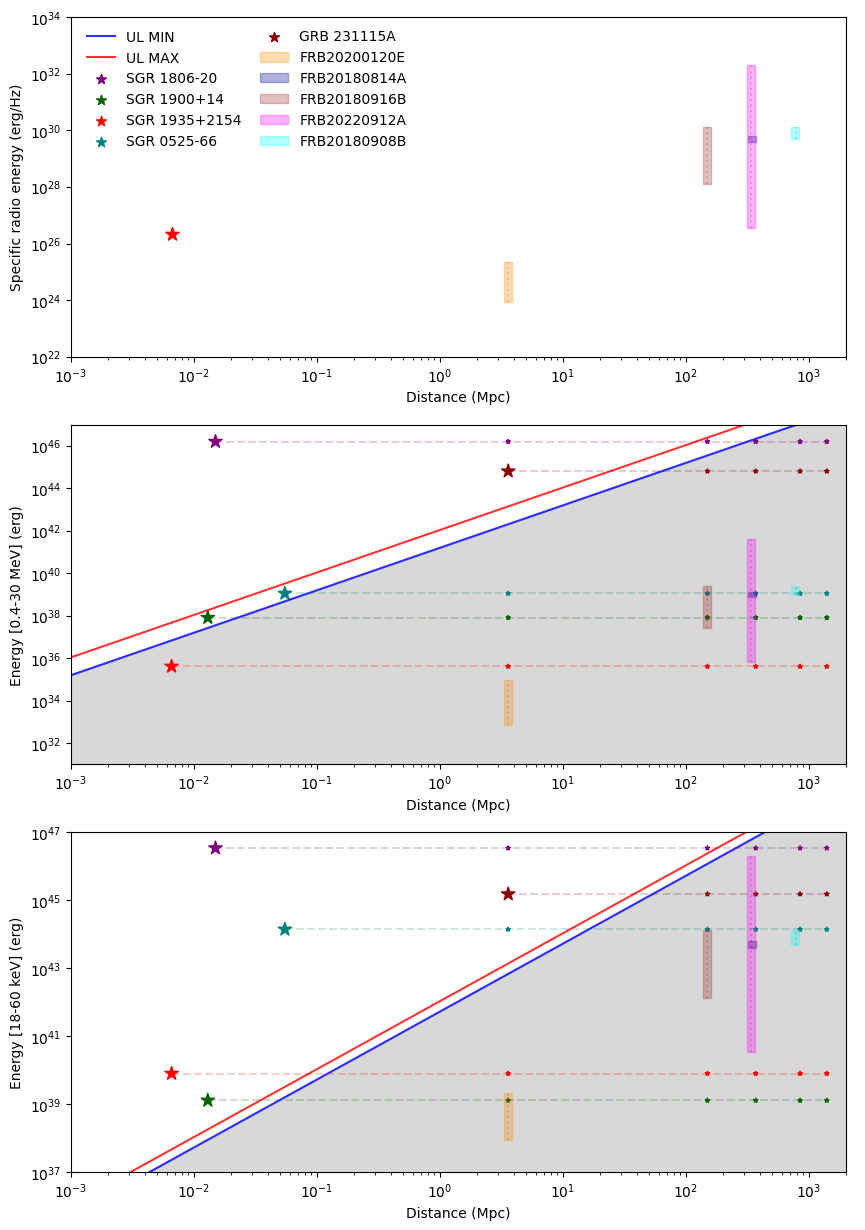}
    }
   \caption{Comparison of radio and high-energy emissions from FRBs and magnetars at their specific distances. \textit{Top panel:} Isotropic radio energy as a function of source distance. The red star represents the radio energy of the burst associated with SGR 1935+2154. The colored strips correspond to the selected R-FRBs for which we have contemporaneous MCAL data. \textit{Middle panel:} Isotropic extrapolated energies in the MCAL band ($0.4-30$ MeV) as a function of distance. This panel includes scaled energies of the selected R-FRBs and the X-ray burst of SGR 1935+2154. Additionally, it shows rescaled X-ray energies of key Galactic and extragalactic magnetars at their actual distances, with energies extended to other distances for comparison (dashed lines). This provides insight into how the expected emission from FRBs compares to the energy released by these magnetars. \textit{Bottom panel:} Isotropic energies in the SuperAGILE band ($18-60$ keV) as a function of distance, illustrating the behaviour of high-energy emissions in this band for the same sources. Middle and bottom panels shows how the \agile sensitivity would allow the detection of SGRs but not of this sample of R-FRBs.}
 \label{Figure_6}
\end{figure}

When simultaneous data acquisition with radio bursts is available, the flux UL derived from MCAL data is substantially lower than previous estimates \citep{Casentini2020, Trudu2022}, and similar to other results in the high-energy bands. Since the temporal width of these bursts is often lacking, we assume a duration of $100 \, \mathrm{ms}$ for our calculations, based on the duration of the most intense component of FRB 200428 \citep{Mereghetti2020} and some known giant bursts from galactic SGR (discussed in the following). This assumption allows us to estimate the energetics reliably. In the absence of a detected signal, we infer that any emission is hidden in the detector's background. Thus, the real astrophysical flux is expected to be smaller than the values listed in the fifth column of Table~\ref{tab:Table_3}. The fluence corresponding to the assumption of burst duration of $100 \, \mathrm{ms}$ is constrained to be below $\sim 10^{-9} \, \mathrm{erg\,cm^{-2}}$, a value which is compatible with those reported in other multiwavelength works \citep[see for instance][]{Trudu2022, Ridnaia2024, Pelliciari2024}. In the case of FRB 20200120E, we obtain a luminosity limit in the range of $\sim 4\times 10^{43} \, \mathrm{erg\,s^{-1}}$, which is lower with respect to the limit imposed by Konus-Wind results with the same FRB 200428 model. Using MCAL's background constraints, we calculate the threshold energy required to trigger the detector at various distances through equation

\begin{equation}
    E_{\mathrm{MeV,UL}} \ [\mathrm{erg}] = 4 \pi d^2 \, {\Delta}t \, F^{\mathrm{UL}}_{0.4-30 \, \mathrm{MeV}}
    \label{eq:MeV_UL}
\end{equation}

where $d$ is the distance of interest, $\Delta t$ is the assumed duration of $100$ ms, and $F$ is the $3 \sigma$ flux UL reported in Table~\ref{tab:Table_3}. We used the highest and lowest UL values in Table ~\ref{tab:Table_3} to evaluate the minimum and maximum energy thresholds that define the sensitivity of MCAL. The threshold spans between $\sim 10^{37} \, \mathrm{erg}$ at $10^{-2} \, \mathrm{Mpc}$ to $\sim 10^{47} \, \mathrm{erg}$ at $10^{3} \, \mathrm{Mpc}$, as depicted by the blue diagonal solid line in second panel of Figure~\ref{Figure_6}. The gray-shaded region below this line corresponds to parameter spaces where detection is infeasible for $100 \, \mathrm{ms}$ bursts. Additionally, the red line in the same panel represents the sensitivity threshold based on the maximum flux UL.

Moreover, we estimate the energy of the five FRBs for which we calculated $3$\textsigma\, ULs within the MCAL energy band ($0.4-30 \, \mathrm{MeV}$), inferred from their radio energetics. \citet{Verrecchia2021} reported an X-ray-to-radio energy ratio of $\sim 10^{5}$ for FRB 200428, the only FRB with an observed X-ray counterpart. This ratio is significantly lower than previous and other estimates, which range up to $\sim 10^{8}$ \citep{Casentini2020,Tavani2021a,Laha2022}, inferred from the magnetar model applied to nearby repeaters. Given the FRB 200428 unique hempirical ratio, we adopt it as a baseline assumption. It is important to note that the energy ratio in \citet{Verrecchia2021} was evaluated referring to the SuperAGILE energy band (the X-ray burst was observed in [$18-60$ keV]). This means that, to evaluate the ratio in the MCAL band, we had to extrapolate the X-ray flux in the [$0.4-30$ MeV] energy range for the cutoff power law model that describe the observed burst \citep{Mereghetti2020}: following this procedure we obtain a ratio $E_{MCAL} / E_{radio} \sim 2.2$ . This very low value is biased by the fact that the X-ray burst observed by that single event is very faint in the X-rays, with respect to the typical X-ray burst fluxes during a SGR activity phase. Furthermore, the energy cut-off of the model is outside the MCAL band (E$_p \sim 65$ keV) which limit the extrapolated flux.

We evaluated the specific energy in the radio band for events with available data at T$_0$ as in Section~\ref{sec-3}, assuming $1$ ms duration and 1 GHz bandwidth for every burst. The results are presented in the top panel of Figure~\ref{Figure_6}. Using the aforementioned energy ratio, we derived the vertical bars representing the energy in the $18-60 \, \mathrm{keV}$ band, shown in the bottom panel of Figure~\ref{Figure_6}. Using the $2.2$ value for the extrapolated hard-X-ray/radio ratio, we derived vertical bars representing the energy in the $0.4-30 \, \mathrm{MeV}$ band, shown in the central panel of Figure~\ref{Figure_6}. In no case, even for the most energetic radio bursts observed, or the closer one, it is expected that they reach the level to trigger MCAL or to be visible for SuperAGILE: except for the most energetic cases. The emission is expected to be (on average) more than an order of magnitude less energetic than what is required to have a signal in our detectors. 

We also include case studies of notable SGR sources. Observational and theoretical evidence increasingly suggests a link between magnetars and at least a subset of FRBs. In particular, the discovery of FRB-like bursts from the Galactic magnetar SGR 1935+2154 has strengthened this connection \citep{Bochenek2020, CHIME2020, Mereghetti2020, Ridnaia2021}. Giant flares and intense bursts from known SGRs share some phenomenological similarities with FRBs, including short durations and high brightness temperatures. These events are hypothesized to originate in the dynamic magnetospheres of magnetars, potentially serving as the progenitors of FRB emission \citep{Margalit2020, Tavani2021b}. Given the established link between at least one FRB (FRB 200428) and a magnetar, we compare the energetics and distances of our FRB sample with known SGR flares, either giant or of medium brightness, to explore whether similar mechanisms could power extragalactic FRBs and to assess the detectability of such events with MCAL.

We focus on giant flares such as those from SGR 1806-20 \citep{1806-20}, GRB231115A in M82 \citep{M82}, and the March 5, 1979 event from SGR 0525-66 in the LMC \citep{March_5th}. Additionally, we examined non-giant bursts from SGR 1900+14 \citep{1979SvAL....5..343M, 2017ApJS..232...17K} and SGR 1935+2154 \citep{2014ATel.6294....1C, 2020ApJ...898L..29M}. For each source, we modeled the spectrum of bursts having a cut-off power-law model to estimate flux and energy in the SuperAGILE and MCAL bands. In MCAL we expect that giant bursts like those from SGR 1806-20 remain detectable up to $\sim 100 \, \mathrm{Mpc}$, whereas non-giant bursts are detectable only within tens of kpc, consistent with the non-detection of SGR 1935+2154 in April 2020 \citep[as reported in][]{Tavani2021b}. In the SuperAGILE band, due to the spectrum of these emissions, we expect higher detectable energies from the same sources, as shown for the lower panel of Figure~\ref{Figure_6}. In this panel we used both the highest and the lowest ULs reported in Table~\ref{tab:Table_4} to produce the detection thresholds for SuperAGILE (as shown by the red and blue diagonal solid line).

We would have the opportunity to observe close FRBs up to few Mpc if their emission mechanism in the hard-X band is related to that of the Giant Bursts historically observed from various SGR. For example, in the case of FRB 20200120E \ (the closer one $@ 3.6 \mathrm{Mpc}$ in M81) we could have observed it if its emission was energetic like GRB231115A. Up to now, its extrapolated energy lies in the energy range of the non-giant SGR activity, which was inaccessible to us at that distance in the hard-X energy band. 

We considered, also in this work, possible \textgamma-ray emission on long timescales, from months to years, which is expected in some models. Considering the $2$\textsigma\,UL on the longest integration of $\sim\,$17 years of $3\, \times 10^{-13}\, -  \, 3\,\times 10^{-12}\,\rm erg\,cm^{-2}\,s^{-1}$ we can infer the expected isotropic luminosity limit at the FRB 20200120\, distance, $L_{\gamma,UL} \simeq  (4-50)\,\times\, 10^{38} \rm \, d_{3.6\,\rm Mpc}^2 \, erg \, s^{-1}$, which puts significant constraints on the non-detection of a signal in \textgamma-ray (in some model, the expected MeV-GeV emission at the SGR 1935+2154 distance should be of the order of $10^{-12}\,\rm erg\,cm^{-2}\,s^{-1}$, \citealt{Metzger2019}) from sources at this distance, but only on integrations above tens of years.

\begin{figure}[ht]
   \centerline{
\includegraphics[width=0.7\textwidth, angle = 0]{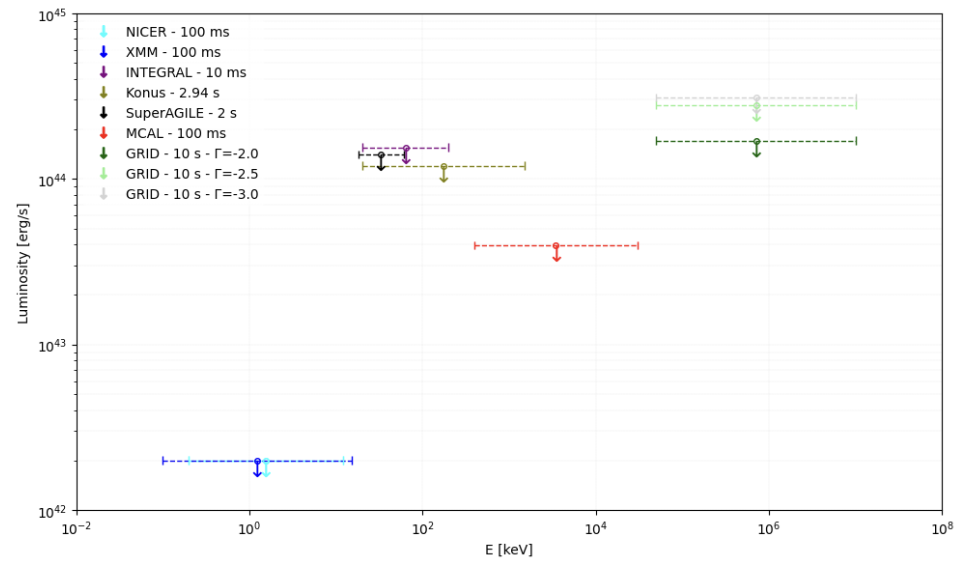}
    }
   \caption{Comparison of the luminosity ULs presented in this work (black point for SuperAGILE, red point for MCAL and green/lightgreen/lightgray points for GRID) with those reported by \citealt{Pearlman2025} (blue and cyan points), \citealt{Ridnaia2024} (olive point) and \citealt{Topina2021} (purple point), even if on different energy bands. Integration times used to evaluate luminosity limit values are reported in the plot legenda. Regarding the GRID observations, we considered here only the UL values for the $10$\,s integration after T$_\mathrm{burst}$ evaluated with a power law model for different spectral indices. We report in darkgreen, and in green/lightgray the associated luminosity UL values for the spectral indices $-2.0, -2.5$ and $-3.0$, respectively. The dashed horizontal lines represent the energy band of each instrument.}
 \label{Figure_7}
\end{figure}

Our results are compatible with those from previous searches for high energy counterparts. In particular, regarding FRB 20200120E, our flux ULs (based on the same cut-off power law spectral model from \citealt{Mereghetti2020}), even if on different energy bands, allow to obtain luminosity limits about a factor 1.5 higher than in \citealt{Pearlman2025} for MCAL, while about a factor of 3 higher for SuperAGILE. However, these last values are of the same order of magnitude of the INTEGRAL ULs in \citealt{Topina2021} and Konus ULs in \citealt{Ridnaia2024}, as shown in Figure~\ref{Figure_7}. Instead, the ULs of MCAL are the most stringent ULs currently available in the $0.4-30$ MeV energy band. The SuperAGILE data analysis was executed on short time scale only, taking into account a single scanning of the source during the spinning mode at the burst time, while studies on longer integrations will be reported in future works. 

\section{\bf{Conclusions}} \label{sec:sec8}
\label{conclusion}

In this work, we presented a search for potential high-energy (MeV-GeV) counterparts for a sample of selected repeating fast radio bursts (R-FRBs) that are currently known. The sample was defined by selecting sources with $\mathrm{DM_{exc}} < 300 \ \mathrm{pc \ cm^{-3}}$. Although $\mathrm{DM_{IGM}}$ could have been used directly, the inclusion of the host galaxy component in the selection parameter was preferred due to the lack of redshift information for some sources, as shown in Table~\ref{tab:Table_1} (see Section~\ref{sec-3}). The selected sample within this $\mathrm{DM_{exc}}$ range includes well-known, highly active sources such as FRB 20200120E (currently the closest observed R-FRB, located at only 3.6 Mpc), FRB 20180916B (the first periodic repeater), FRB 20201124A, and FRB 20220912A, which have 2883 and 1529 published bursts to date, respectively, also thanks to observations by the FAST radio telescope which is sensitive to very low-fluence events (up to $10^{-3} \mathrm{Jy}\cdot\mathrm{ms}$). An additional criterion was applied: we excluded sources without topocentric time information provided by the experiments. This refinement resulted in a sample comprising bursts predominantly observed by the CHIME radio telescope.

After selecting the R-FRBs of interest, we analyzed them using the detectors onboard the \agile satellite: MCAL, GRID and SuperAGILE. For them, we first verified the visibility of the bursts from the selected sources. Out of the sub-sample containing 747 bursts from 21 sources, we found that $79\%$ were fully within MCAL's visibility window and $25\%$ within GRID's visibility window, while just six bursts were exposed by the fully coded part of the SuperAGILE FoV. For MCAL, we identified $25$ bursts that not only fell within its visibility but were also temporally coincident with its data acquisition triggers.

Upon analyzing them, no evidence of astrophysical signals was found in temporal coincidence\,(within 1\,s around burst times). The triggers were attributed to charged particles interacting with the detector or random coincidences from background noise. The individual trigger analysis yielded a False Alarm Rate (FAR) of approximately 100 events per day, consistent with background fluctuations. Finally, we estimated the minimum flux required for a $3\sigma$ detection using a CPL spectral model.

We further calculated the energy thresholds and derived observational constraints on FRB energetics within the MCAL energy band ($0.4-30\, \mathrm{MeV}$), focusing on five sources with $3\sigma$ flux upper limits. These calculations, assuming isotropic emission, show that even the most energetic FRBs observed are unlikely to trigger MCAL, except under extreme conditions involving mechanisms analogous to giant flares from magnetars. Additionally, we analyzed notable SGR sources and determined that while giant bursts could be detectable up to $\sim 30 \ \mathrm{Mpc}$, non-giant bursts remain observable only within tens of kpc. For nearby FRBs, such as FRB 20200120E at $3.6$ Mpc (currently the nearest localized FRB source), the energetics are consistent with non-giant SGR activity, which is inaccessible at such distances with our current detection capabilities.
Furthermore, we repeated the check on \textgamma-ray emission on long time scales, at increased exposure with respect to \cite{Verrecchia2021}, obtaining a significantly low luminosity limit at $\sim\,10^{39}\,\rm d_{3.6\,\rm Mpc}^2 \, erg\,s^{-1}.$

The \agile mission, that ended in $2024$ with the satellite’s re-entry into the atmosphere, has provided a wealth of data for the study of FRBs. While new observations with \agile are no longer possible, the extensive archival data offer significant opportunities for further analysis, also for non-repeating FRBs. These data can help refine our understanding of FRB energetics and their high-energy counterparts, potentially uncovering new insights into their origins and emission mechanisms. Future missions with advanced sensitivity and broader energy coverage will be essential to build on \agile’s legacy and further discover the multi-wavelength nature of FRBs.\\

{\bf Acknowledgments:} Investigation carried out with partial support by the ASI grant no.\,I/028/12/05 for the AGILE project. We would like to acknowledge the financial support of ASI under contract to INAF: ASI 2014-049-R.0 to ASI-SSDC. Claudio Casentini dedicates this work to his late mother Carla.


\bibliographystyle{aasjournal}

\end{document}